# Structural Properties of of Group IV B Element Rutherfordium by First Principles Theory

Jyoti Gyanchandani[1] and S.K.Sikka[2]

[1] Material Science Division, Bhabha Atomic Research Center, Mumbai 400085, India

[2] Bhabha Atomic Research Center, Mumbai 400085, India

The structural and electronic properties of Rutherfordium, the newest group IV B element, have been evaluated by first principles density functional theory in scalar relativistic formalism with and without spin-orbit coupling and compared with the known experimental and theoretical properties of its homologue Hf. It is found that it will crystallize in the hexagonal close packed structure with metallic character. However, under pressure, it will have different sequence of phase transitions than Hf: hcp→bcc instead of hcp→ω→bcc.

PACS numbers: 71.15.Nc, 64.30.-1, 71.30.Be

Keywords: Rutherfordium, DFT, cohesive properties, pressure

## 1. Introduction

Rutherfordium (Rf), the 104[th] element of the periodic table, was first synthesized in 1964 at Dubna by bombarding Pu with accelerated Ne ions [1]. Subsequently in 1969, it was also discovered at Berkeley by bombarding a target of Cf with beams of carbon isotopes [2]. In 1970, Ghiorso et al [3] found a new isotope, $Rf^{261}$ with a half life of 69s. Later on, it has been established that some isotopes of Rf have much longer half lives [4] ( $Rf^{265}$ ~ 13 hrs and $Rf^{267}$ ~ 1 hrs ). These relatively long half-life enabled the study of the chemistry of this element and the experimental results showed that it was a group IV element like tetravalent Zr and Hf [5]. The results of the relativistic atomic calculations also revealed its ground state electronic configuration to be $6d^2 7s^2$, analogues to lighter group four elements[6]. Because of the paucity of the available amounts of this material for investigation of solid state properties, none of the latter have been measured. We have, instead, employed first principles density functional theory to fill this gap. Due to the strong relativistic effects expected here, significant deviations from periodic trends may be probable. Such theoretical calculations are beginning to be done for trans-actinides [7,8,9} and have already revealed some surprises. For example, element Cn (eka Hg), a homologue of Hg, is most likely to be a semiconductor instead of a metal. We have also just published a brief account of the results on all the 6d series of elements [10]. In particular, we have determined the 0K crystal structures and equation of state parameters It is demonstrated that these cohesive properties of this series have similar trends as the 5d series. Friedel theory [11] used to explain these trends can also be applied for 6d metals. Here, we give more details for Rf. For comparison, we have also performed calculations for Hf.

## 2. Computational Method

First principles structure optimizations were done through total energy computations at 0 K by the full potential linearized augmented plane wave (FP-LAPW) method as implemented in WIEN2K code [12,13]. All the calculations were carried out on Anupum Ajeya supercomputer at Bhabha Atomic Research Centre. We utilized the Perdew Burke Ernzerof generalized gradient approximation (GGA) for exchange and correlation functional, incorporating scalar-relativistic effects (SR) [14]. Calculations were carried out with and without including spin orbit (SO) coupling to find out its effect. This is because for heavy elements SO effects have been shown to be important as these scale roughly as $Z^2$ [6]. For example, for the element Uuq (Z=114), Hermann et al [9] recently showed that its equilibrium crystal structure was hexagonal close packed (hcp) in SR+SO calculations compared to face centred cubic (fcc) in SR ones. The 6d, 7s and 7p electrons were included as conduction states while the 5f, 6s and 6p electrons were treated as semi-core states. The computations have been done for hcp, bcc, fcc and ω structures. The ω structure is of $AlB_2$ type in space group P6/mmm, having 3 atoms in its unit cell located at 0,0,0, 1/3,2/3,1/2 and 2/3,1/3,1/2. It may be noted that this structure occurs for Ti, Zr and Hf under high pressure [15]. A dense grid of 5000 k points was employed for sampling the Brillouin zones in order to avoid any effects of dissimilar shapes of these for different structures. . The plane wave cutoff parameter $R_{MT} K_{MAX}$ was chosen as 9 with muffin tin radius $R_{MT}$ to be 2.4 a.u. The self consistent cycle in each case was run until the energy convergence criterion of 0.01 mRy was met.

## 3. Results

Table 1 lists the derived parameters: lattice constants, nearest neighbour distance ($d_0$), bulk modulus ($K_0$) and its pressure deivative ($K'_0$) for different structures. These parameters were derived by fitting the total energy to Birch-Murnagham equation of state. The total energy versus volume plots (Fig.1) clearly show that the hcp structure is the most stable structure for Rf at normal volume. The same is true for Hf. For it, the minimum energy structure determined in this investigation and the calculated parameters are in excellent agreement with the experimental data. There is hardly any difference between SR and SR+SO results. This is in accord with the findings of Fang et al [16]. However, for Rf., the lattice parameters for different structures are slightly shrunk (< 1%) when SO coupling is included. But, the overall effect is small.

To ascertain the other possible high pressure phases of Rf, we have computed the zero-temperature enthalpies, G = E+PV for different candidate structures as a function of pressure. This analysis points out that Rf will transform from hcp to the bcc phase at about 72 GPa in SR and ~50 GPa in SR+SO treatments (Fig.2). In this respect, Rf differs from Hf. High pressure experiments on Hf by Xia et al [17] found hcp → ω phase change at 38±8 GPa and ω → bcc transition at 71±1 GPa. In our Gibbs energy evaluations for Hf, these occur around 44 and 72 GPa respectively. It may be argued that the disagreement between Rf and Hf may be attributed to thermal effects as the experimental data of Hf is at room temperature. However, a simple estimation of thermal energy ($E_{therm} \approx 3kT$) and thermal pressure ($P_{therm} \approx \gamma/V\ E_{therm}$ with $\gamma = 2\ K_0' -1$) shows that this is not so at T=300 k. Fig 2 presents the density of states of hcp Rf for both SR and SR+SO cases. It may be noted that there is hardly any difference between the two cases in the region below the Fermi energy. These also reveal that Rf is a metal with predominant d character. Band structure plots (not given here) show that the 5f states are located about 13 electron volts below the Fermi level.

The electronic configuration of the conduction states of Rf in the solid state at equilibrium volume is found to be ~ $6d^{1.77}7s^{0.7}7p^{0.38}$. This changes to ~ $6d^{1.97}7s^{0.54}7p^{0.29}$ corresponding to the hcp to bcc transition pressure. This is in conformity with the prevalent view that high pressure transformations in group IV solids are driven by s→d electron transfer [18]. However, no quantitative reason can be attributed for absence of hcp → ω transiton in Rf from density of states or change in band occupancies. Ahuja et al [19] have plotted canonical one electron energies as a function of d band filling in transition metal series and show that ω structure is favoured at d-fillings close to 2 and 6-8. However, this is opposed by the Madelung contribution, which stabilizes the more close packed structures. May be this competition between two contributions is responsible for the absence of the relatively open ω phase with apacking fraction of 0.57 compared to 0.68 for the bcc structure in Rf under pressure.

The atomic volume of Rf is about 10% larger than Hf. This may be compared with 10% for Cn (Z=112) with respect to Hg [7] and 15% for fcc Uuq(Z=114) to Pb[9]. This expansion is also in line with that found in the geometry optimization studies of various gas phase chemical compounds of Rf. There, this has been ascribed to enhanced relatvistic effects in heavy elements and in particular due to both the orbital and relativistic expansion of the 6d orbitals, compared to the 4d and 5d ones [20]. The atomic weight of Rf is listed as 262.11 [21]. From this value, the expected density of Rf is found to be 17.9 gm/c.c. This is about 35% higher than Hf. It is well established that the minimum distance in the ω phase (bond length between the two atoms situated on the c/2 plane) represents twice the Pauling's univalent radius. This gives the covalent radius of Rf as 1.485 Å. The value 1.442 Å for Hf matches the value listed by Pauling [22]. However, both the values for Rf and Hf are lower than the values 1.57 and 1.52 Å estimated recently by Pyykkö and Atsumi [23]. The theoretical bulk modulus and its pressure derivative of Rf fall into the parabolic and linear trends respectively exhibited by the 6d solids very similar to those of 4d and 5d series of transition metals( see figures 2-4 in [10]).

In conclusion, our DFT calculations show that with and without SO coupling there is no significant effect on the values of the structural parameters of Rf. Its atomic volume is expanded with respect to its 5d homologue Hf due to stronger relativistic effects. Under pressure, it will undergo hcp-bcc phase change instead of hcp-ω-bcc in Hf

JG thanks Dr.G.K.Dey, Head Materials Science Division,BARC for constant encouragement.

**References**

1. Flerov G N, Oranesyan Yu Ts., Lobanov Yu V, Kuznetsov V I, Druin V A, Perelygin V P, Gavrilov K A, Tretiakova S P and Plotko V M 1964 Phys. Lett.13 73
2. Ghiorso A, Nurmia N, Harris J, Eskola K and Eskola P 1969 Phys.Rev.Lett. 22 1317
3  Ghiorso A, Nurmia N, Eskola K and Eskola P  1970 Phys.Lett. B32 95
4. Holden N E *Radioactive elements in the standard atomic weights table* 2007 BNL-79969
5. Pershina V *The Chemistry of Superheavy Elements*, edited by Schädel M, 2003 (Kluwer Academic Publishers, Dordrecht,) p 31
6. Pershina V *Relativistic Electronic Structure Theory*, Part 2, edited by P. Schwerdtfeger 2004 ( Elsevier, New York, ) p.1
7. Gaston N, Opalhle I, Gäggerer H W and Schwerdtfeger P, 2007 Angew. Intl. Ed. 46 1663
8. Noffsinger J and Cohen M L 2010 Phys.Rev. B81 073110
9. Hermann A, Furthmüller J, Gäggeler H W  and Schwerdtfeger P 2010 Phys. Rev. B82, 155116
10. Gyanchandani Jyoti and Sikka S K 2011  Phys. Rev. B83 172101
11. Friedel J, *The Physics of Metals: Electrons* edited by J.M.Ziman 1969  (Cambridge Uni. Press ) p.340
12. Blaha P, Schwarz K and  Luitz J  *WIEN2K*  2010 (Technical University of Vienna)
13. Kunes J, Novak P,  Schmid R, Blaha P and Schwarz K 2001 Phys. Rev  B 64 153102
14. Perdew JP , Burke K  and Ernzerhof  M , 1996 Phys. Rev. Lett. **77**, 3865
15. Sikka S K, Vohra Y K and Chidambaram R  1982 Prog. Mater. Sci. 27 245

16. Fang H, Gu M.Liu B, Liu X, Huang S, Ni C, Li Z and W R 2011 Phyica B 406 1744
17. Xia Hui, Parthasarathy G, Luo Haun, Vohra Y K and Ruoff A L 1990 Phys.Rev. B42 6736
18. Vohra Y K 1979 Acta Metall. 27 1671
19. Ahuja R , Wills J B, Johansson B and Eriksson O 1993 Phys. Rev.B 48 16269
20 Pershina V 2009 Russ. Chem Rev. 78 1153 and references there in
21. Physical Measurement Laboratory of The National Institute of Standards Technology (NIST), an agency of U.S Commerce Department ([www.nist.gov/pml](www.nist.gov/pml) /data/comp.cfm)
22 Pauling L *The Nature of the Chemical Bond* 1968 3$^{rd}$. ed., Cornell University Press. Ithaca, NY
23. Pyykko P and Atsumi M 2009 Chem.Eur.J. 15 186
24 Kittel C *Introduction to Solid State Physics* 1971 Wiley, New York
25 Donohue J *The Structures of the Elements* 1974 Wiley, New York
26. Steinberg D J 1982 J.Phys.Chem Solids 43 1173
27. Tonkov E Y and . Ponyatovsky E G 2005 *Phase Transformations of Elements Under High Pressure*, CRC Press, Boca Raton, F28.

Table1. Lattice constant (Å), c/a ratio if any, interatomic distance $d_0$ (Å), bulk modulus $K_0$(GPa) and pressure derivative of bulk modulus $K_0'$ at equilibrium volume for various candidate structures for Rf

|  |  | Theory (SR) | | | | | Theory (SR+SO) | | | | | Experiment | | | | |
|---|---|---|---|---|---|---|---|---|---|---|---|---|---|---|---|---|
| Hf | Phase | $a_o$ | $(c/a)_o$ | $d_o$ | $K_o$ | $K_0'$ | $a_o$ | $(c/a)_o$ | $d_o$ | $K_o$ | $K_o'$ | $a_o$ | $(c/a)_o$ | $d_o$ | $K_o$ | $K_o'$ |
|  | hcp | 3.194 | 1.581 | 3.127, | 107 | 3.44 | 3.201 | 1.581 | 3.133 | 108 | 3.41 | 3.190[a] | 1.583[a] | 3.127 | 109[a] | 3.96[b] 3.44 |
|  | ω | 4.996 | 0.620 | 2.884 3.097 | 107 | 3.61 | 4.993 | 0.620 | 2.882 3.095 | 107 | 3.64 | 4.947[c] | 0.622[c] | 2.856 3.079 | | |
|  | bcc | 3.545 | | 3.071 | 99 | 3.36 | 3.543 | | 3.068 | 96 | 3.56 | 3.615[d] | | 3.131 | | |
|  | fcc | 4.485 | | 3.171 | 101 | 3.44 | 4.470 | | 3.160 | 100 | 3.55 | | | | | |
|  |  |  |  |  |  |  |  |  |  |  |  |  |  |  |  |  |
| Rf | hcp | 3.302 | 1.575 | 3.226 | 94 | 3.84 | 3.269 | 1.590 | 3.212 | 100 | 3.69 | | | | | |
|  | ω | 5.135 | 0.627 | 2.969 3.220 | 94. | 3.97 | 5.112 | 0.620 | 2.951 3.135 | 96 | 4.00 | | | | | |
|  | bcc | 3.643 | - | 3.155 | 88 | 4.03 | 3.608 | | 3.125 | 93 | 4.11 | | | | | |
|  | fcc | 4.589 | - | 3.245 | 91 | 3.87 | 4.578 | | 3.237 | 93 | 3.94 | | | | | |

a from reference [24,25] ; b from reference [26] ; c from reference [15] ; d from reference [27] at T > 2053

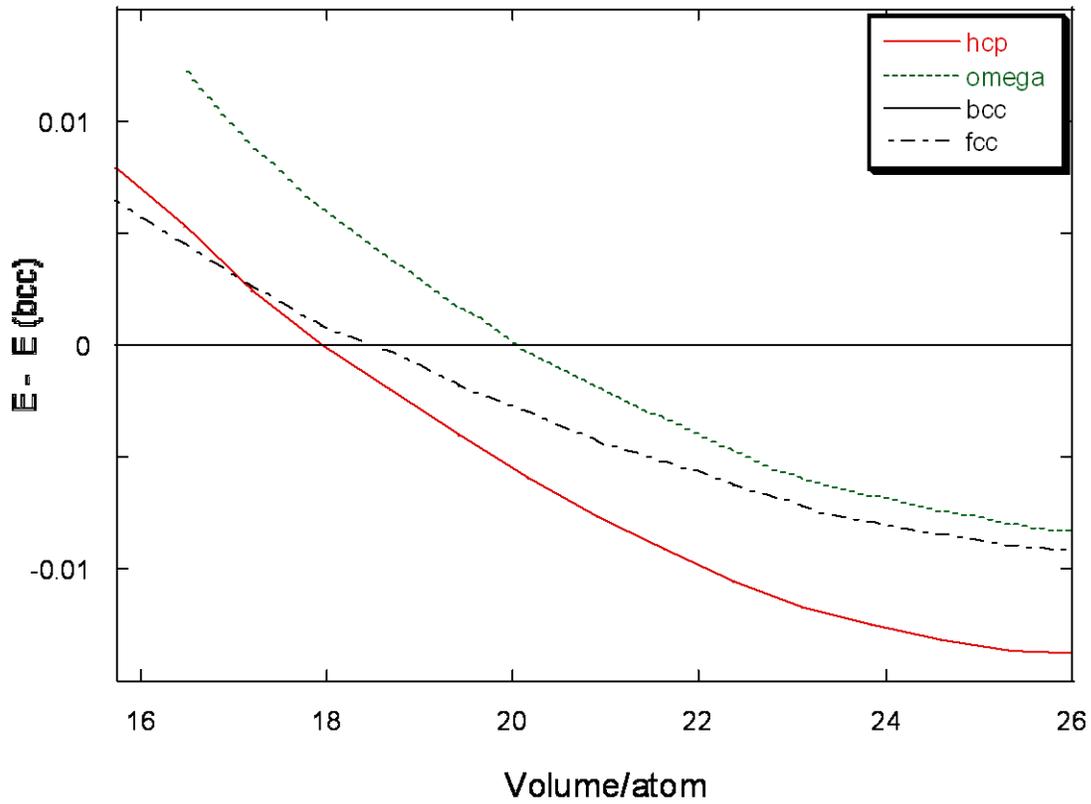

Fig.1　Energy differences in Rydberg with respect to the bcc structure versus atomic volume ($Å^3$) for various structures of Rf for SO coupling case.

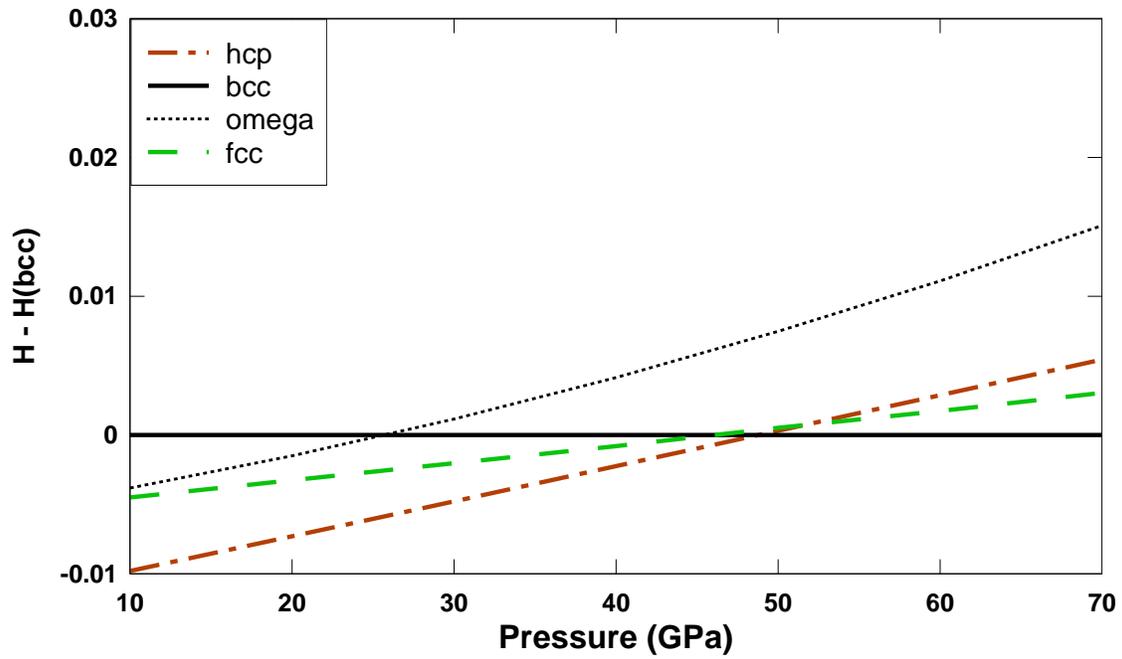

Fig.2_Enthalpy differences in Rydberg with respect to the bcc structure versus pressure for various phases of Rf for SO coupling case.

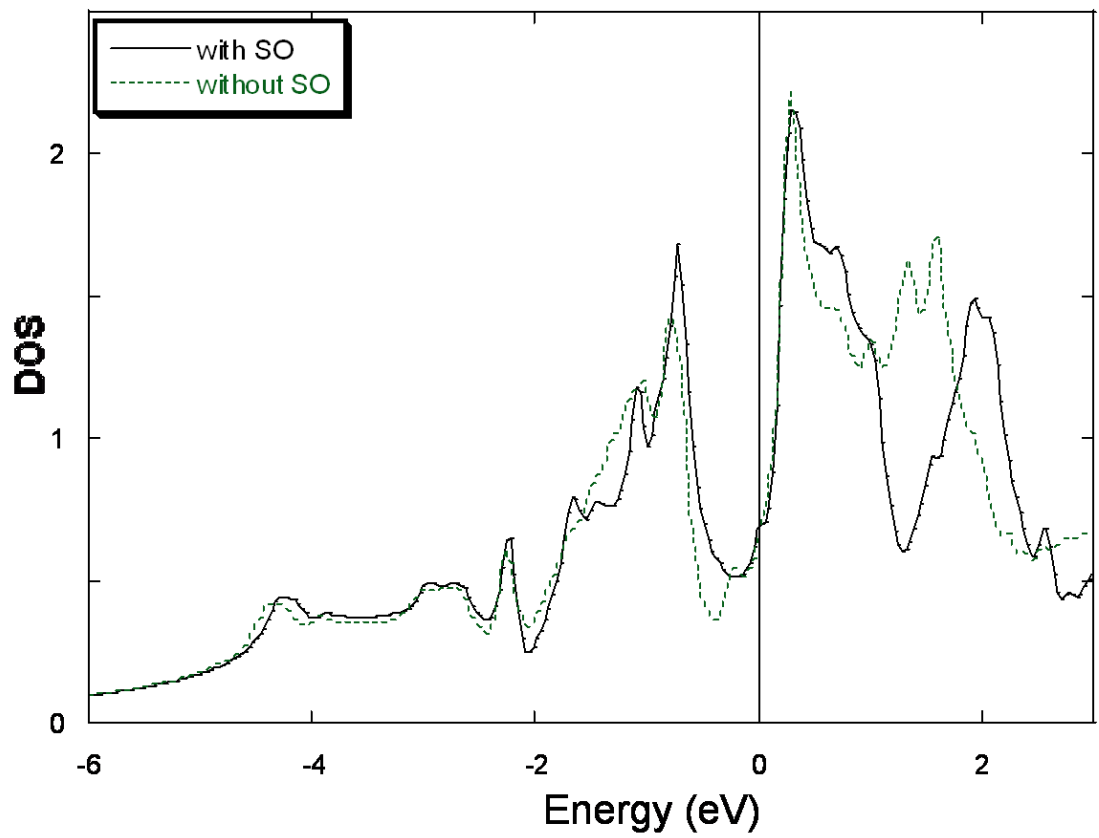

Fig.3. Total density of states (DOS) for Rf at equilibrium volume with and without SO